\documentclass[doublecol]{epl2}

\title{Crossover between different universality classes: Scaling for thermal transport in one dimension}

\author{Daxing Xiong \thanks{E-mail: \email{phyxiongdx@fzu.edu.cn}}}

\institute{
 Department of Physics, Fuzhou University, Fuzhou 350108, Fujian, China
}


\abstract{For thermal transport in one-dimensional (1D) systems,
recent studies have suggested that employing different theoretical
models and different numerical simulations under different system's
parameter regimes might lead to different universality classes of
the scaling exponents. In order to well understand the universality
class(es), here we perform a direct dynamics simulation for two
archetype 1D oscillator systems with quite different phonon
dispersions under various system's parameters and find that there is
a crossover between the different universality classes. We show that
by varying anharmonicity and temperatures, the space-time scaling
exponents for the systems with different dispersions can be feasibly
tuned in different ways. The underlying picture is suggested to be
understood by phonons performing various kinds of continuous-time
random walks (in most cases, be the L\'{e}vy walks but not always),
probably induced by the peculiar phonon dispersions along with
nonlinearity. The results and suggested mechanisms may provide
insights into controlling the transport of heat in some 1D
materials.}

\begin{document}

\maketitle






Transport in one dimension has, for a long time, been realized to be
anomalous in most cases~\cite{Anomalous-1,Anomalous-2}, with
signatures of a universal power-law scaling of transport
coefficients, among which the heat transport has been extensively
investigated in the recent decades, both by various theoretical
techniques, such as the renormalization group~\cite{Renormalize},
mode coupling~\cite{MCT-1,MCT-2} or
cascade~\cite{Lee-Dads-1,Lee-Dads-2,Nonuniversal_new}, nonlinear
fluctuating
hydrodynamics~\cite{Hydrodynamics-1,Hydrodynamics-2,Hydrodynamics-3},
and L\'{e}vy walks~\cite{Levy-1,Levy-2,Levy-3,Levy-4}; and also by
computer
simulations~\cite{Grassberger,Numerics-1,Numerics-2,Numerics-3,Numerics-4,Numerics-5,Numerics-6,Numerics-7,Method-1,MyWork,Nonuniversal_new_1}.
For all studied cases two main scaling exponents have been given the
most focus, i.e., $\alpha$ describing the divergence of heat
conductivity with space size $L$ as $L^\alpha$ and $\gamma$
characterizing the space($x$)-time($t$) scaling of heat spreading
density $\rho (x,t)$ as ${t^{-1/\gamma}} \rho (t^{-1/\gamma} x,t)$.
Unfortunately, however, depending on the focused system's different
parameter regimes, different theoretical models have been employed,
and different predictions have been suggested. Thus, the
universality classes of both scaling exponents and their
relationship~\cite{Note,Nonuniversal_new_2} remain controversial:
(i) for $\alpha$, two classes:
$\alpha=1/3$~\cite{Renormalize,Hydrodynamics-1,Levy-3,Levy-4,Grassberger,Numerics-3}
and  $\alpha=1/2$~\cite{MCT-1,MCT-2,Hydrodynamics-1} have been
reported; however, the universality has been doubted~\cite{MyWork,
Nonuniversal_new_1} and a Fibonacci sequence of $\alpha$ values
converging on $\alpha^{*}=(3-\sqrt{5})/2$ ($\simeq
0.382$)~\cite{Nonuniversal_new,Lee-Dads-1,Lee-Dads-2} has been
suggested; (ii) for $\gamma$, two universality classes,
$\gamma=5/3$~\cite{Hydrodynamics-1,Hydrodynamics-2,Hydrodynamics-3,Levy-2,Levy-3,Levy-4,Numerics-4,Numerics-5,Numerics-6,Numerics-7}
and
$\gamma=3/2$~\cite{Hydrodynamics-1,Hydrodynamics-2,Hydrodynamics-3,Numerics-6,Numerics-7}
have been predicted recently.

The discussion of the latter scaling exponent $\gamma$ is currently
very
hot~\cite{Hydrodynamics-1,Hydrodynamics-2,Hydrodynamics-3,Levy-1,Levy-2,Levy-3,Levy-4,Numerics-4,Numerics-5,Numerics-6,Numerics-7,Rotor-1,Rotor-2}
because it involves more detailed space and time
information~\cite{Workshop}, thus it can present a very detailed
prediction for heat transport. It is also relevant to the dynamical
exponents in general transport processes far away from equilibrium,
such as that described by the famous Kardar-Parisi-Zhang (KPZ)
class~\cite{KPZ}. Recently, a new class of dynamical exponent
$5/3$~\cite{PRL2014} and a novel Fibonacci family of dynamical
universality classes~\cite{2015New} beyond the KPZ class, quite
similar to the universality of $\alpha$ and
$\gamma$~\cite{Lee-Dads-1,Lee-Dads-2,Nonuniversal_new}, have been
proposed.

Nevertheless, simulations to precisely identify the dynamical
exponents, especially the exponent $\gamma$ for heat transport, from
direct dynamics, are actually hard to carry out, causing quite few
numerical results
reliable~\cite{Numerics-6,Numerics-7,Rotor-1,Rotor-2}. With this
question in mind, in this work, by employing a direct dynamic
simulation method~\cite{Method-2} we provide a very precise estimate
of $\gamma$ from the new perspective of different phonon dispersions
and under various system's parameter regimes, from which a crossover
between the different universality classes of $\gamma$ can be
clearly identified. We argue that different dispersions along with
nonlinearity could result in quite different microscopic
environments surrounding phonons continuous-time random walking
(CTRW), which then leads to the different scaling exponents. The
results present the very convincing numerical evidences for
identifying the universality classes of $\gamma$ and present also a
suggestive picture for why.

More specifically, to illustrate our viewpoint, two archetype
oscillator systems with and without usual phonon dispersions, i.e.,
a Fermi-Pasta-Ulam-$\beta$ (FPU-$\beta$) lattice and its extension
to bounded double-well (DW) version, will be employed. To identify
the scaling exponent $\gamma$, the heat, rather than the total
energy's fluctuation relaxation~\cite{Method-2} of the systems will
be investigated. We assume that the dispersion may have its effect
along with system's anharmonicity and temperatures. Thus we shall
study how the scaling depends on these parameters.

Both focused systems are momentum-conserved and with an even
symmetric potential, thus appearing to follow the $\gamma=3/2$
universality class predicted by some related theoretical models
under appropriate parameter
regimes~\cite{Hydrodynamics-3,Numerics-6,Numerics-7}, whose
Hamiltonian can be represented by $H= \sum_{k} ^{L} p_{k} ^2 /2 +V
(q_{k+1}-q_{k})$~\cite{Note-1}, where $q_k$ is the displacement of
the $k$-th particle from its equilibrium position, $p_k$ the
momentum, $V (\xi) = \xi^2/2 + \beta \xi^4 /4$ $(-\xi^2/2+\xi^4/4)$
for the FPU-$\beta$ (DW) systems. From the Hamiltonian, it is clear
that the former bears the usual phonon dispersions under harmonic
approximation; however, the latter not. We expect that such key
distinction may affect the universality class(es) of $\gamma$.

The direct dynamics simulation method~\cite{Method-2} adopted here
employs the following spatiotemporal function: $\rho
(x,t)=\frac{\langle \Delta Q_{j}(t) \Delta Q_{i}(0) \rangle}{\langle
\Delta Q_{i}(0) \Delta Q_{i}(0) \rangle}$ to characterize the heat
spreading density, where $\langle \cdot \rangle$ represents the
spatiotemporal average, $\Delta Q_{i}(t)\equiv Q_i(t)- \langle Q_i
\rangle$; $Q_i(t) \equiv \sum Q(x,t)$ is the total heat energy
density in an equal and appropriate lattice bin [the number of
particles in each bin is equal to $n=L/b$ ($n \equiv 2$) with $b$
the total bins number], $Q(x,t)\equiv E(x,t)-\frac{(\langle E
\rangle +\langle F \rangle)M(x,t)}{\langle M \rangle}$~\cite{Liquid}
is the single-particle's heat energy at a absolute displacement $x$
and time $t$, with $E(x,t)$ and $M(x,t)$ the corresponding energy
and mass density, $\langle E \rangle$, $\langle M \rangle$ and
$\langle F \rangle$ ($\equiv 0$) the spatiotemporal average of
$E(x,t)$, $M(x,t)$ and the internal pressure, respectively.

From this definition, the key point here is concerning the particles
heat energy, rather than the usually considered total energy
fluctuations~\cite{Liquid}, which has been verified to be more
directly related to heat conduction~\cite{New_shunda}. In addition,
to calculate the heat spreading, the space variable should be the
absolute displacement $x$ but not the label $k$ of the particle as
conventionally considered in the previous
simulations~\cite{New_shunda}. Now if the density $\rho (x,t)$ has
been obtained and its coincidence to the L\'{e}vy walks profile has
been verified, then a scaling analysis of its central
part~\cite{Levy-1}:
\begin{equation}
\label{eq.1}
\rho (x,t) \simeq \frac{1}{t^{1/\gamma}} \rho
(\frac{x}{t^{1/\gamma}},t),
\end{equation}
identifies the scaling exponent $\gamma$~\cite{New-Note-2}, which
just gives the dynamical exponent for the particular heat transport
without performing complicated theoretical calculations.
\begin{figure}
\centerline{\includegraphics[width=9.cm]{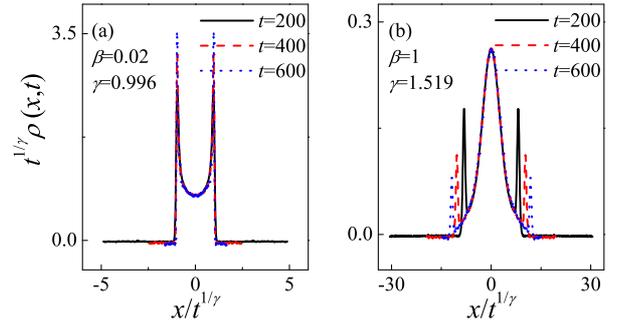}} \vskip-0.3cm
\caption{(Color online) Rescaled profiles of $\rho(x,t)$ for
FPU-$\beta$ chains (with effective space size
$L_{\rm{effetive}}=2000$). In each curve the scaling formula (1) is
applied.} \label{fig.1}
\end{figure}
\begin{figure}
\centerline{\includegraphics[width=9.cm]{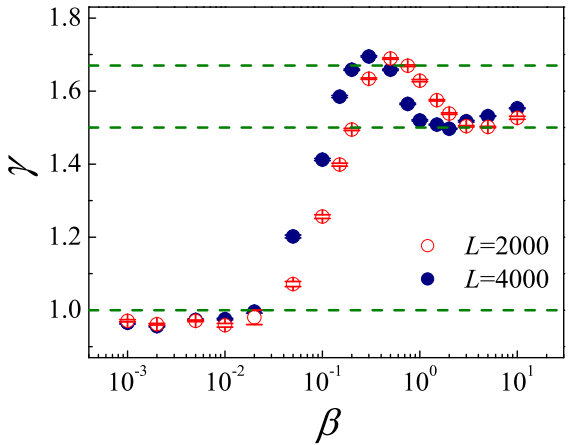}} \vskip-0.3cm
\caption{(Color online) $\gamma$ vs $\beta$ for FPU-$\beta$ chains,
where from bottom to top the dashed lines indicate $\gamma=1$,
$\gamma=3/2$ and $\gamma=5/3$.}
\end{figure}

We first consider the FPU-$\beta$ system. Figure 1 depicts the
rescaled $\rho(x,t)$ for two $\beta$ values, where the temperature
$T=0.5$ is considered~\cite{Note-3}. In both cases the formula (1)
is beautifully satisfied suggesting that their profiles can be well
captured by the L\'{e}vy walks types; however, the scaling is
obviously different. For the frequently considered $\beta=1$ case,
the best fitting gives $\gamma=1.519$ [see Fig. 1(b)], very close to
$\gamma=3/2$ universality
class~\cite{Hydrodynamics-2,Hydrodynamics-3,Numerics-6,Numerics-7};
however, for $\beta=0.02$, with so slight nonlinearity, the best
fitting suggests $\gamma=0.996$ [see Fig. 1(a)], which coincides
well with the $U$-like L\'{e}vy walks propagator in the ballistic
regime~\cite{Levy-1,Levy-Ballistic}, thus suggesting that the
ballistic L\'{e}vy walks theory might be appropriate for describing
ballistic heat transport. It is also interesting to note that this
$U$-shaped distribution has also been found useful in the
statistical description of blinking quantum dots~\cite{Barkai}.

\begin{figure}
\centerline{\includegraphics[width=9.cm]{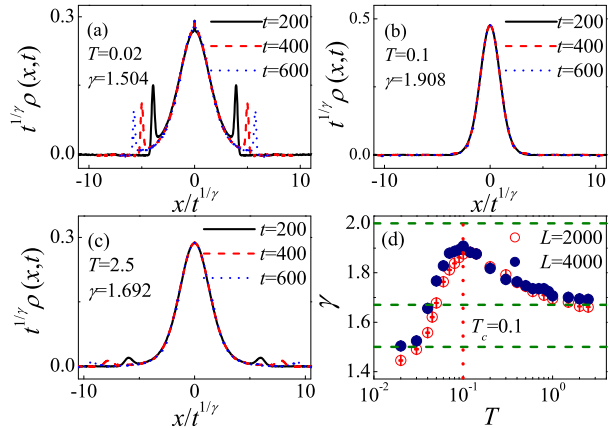}} \vskip-0.3cm
\caption{(Color online) (a)-(c): Rescaled $\rho(x,t)$ for DW chains
($L_{\rm{effetive}}=2000$); (d): $\gamma$ vs $T$, where the dashed
lines, from bottom to top, indicate $\gamma=3/2$, $\gamma=5/3$ and
$\gamma=2$; the dotted one $T_{c}=0.1$; the open (solid) circles
represent the results of $L_{\rm{effetive}}=1000$ $(2000)$.}
\end{figure}

We thus check pictures for other $\beta$ values and summarize them
in Fig. 2, from which, regardless of $L$ and as $\beta$ increases,
$\gamma$ first remains constant at about $\gamma \simeq 1$ ($\beta
\leq 0.02$), then it increases continuously and nonmonotonously. In
particular, for $\beta$ close and up to $1$, most of $\gamma$
concentrate between $3/2$ and $5/3$, then both previous expectations
of
$\gamma=3/2$~\cite{Hydrodynamics-1,Hydrodynamics-2,Hydrodynamics-3,Numerics-6,Numerics-7}
and
$\gamma=5/3$~\cite{Levy-2,Levy-3,Levy-4,Numerics-4,Numerics-5,Method-1}
classes are understandable, because most of them are focused on the
$\beta=1$ case (under appropriate temperatures). Obviously, to more
precisely characterize $\gamma$ in this range, longer space size
simulations are still required~\cite{Note-5}. But anyway, based on
the focused space range here, one can clearly identify a crossover
from ballistic ($\gamma=1$) to superdiffusive ($\gamma > 1$)
transport at about $\beta_c \simeq 0.01$-$0.02$, suggesting that a
finite $\beta$ phase transition would probably take place. Perhaps
more surprised is the continuous variation of $\gamma$ ($1 \leq
\gamma <1.7$), which clearly suggests that there is a crossover
between the different universality classes of $\gamma$.

Next we refer to the DW case. Figure 3(a)-(c) plot the rescaled
$\rho(x,t)$ and (d) presents $\gamma (T)$~\cite{New-Note}. From
(a)-(c) a perfect coincidence with L\'{e}vy walks scaling following
formula (1) (but with different $\gamma$) can be clearly identified.
In addition to this similarity to Fig. 1 and 2, it is worth noting
that around $T=0.1$, in Fig. 3(b) the two side sound modes seem
completely disappeared, causing $\rho(x,t)$ to be very close to the
Gaussian profile ($\gamma=2$). So it is reasonable to see normal
heat conduction ($\alpha=0$) found previously just near this
temperature~\cite{DW-234}. However it should be emphasized that: the
appearance of a diffusive mode shown here seems to indicate that the
two sound modes could be completely absent in some 1D Hamiltonian
systems under appropriate anharmonicity or
temperature~\cite{PRL2014} from the perspective of hydrodynamics,
which may be attributed to the no-bear usual phonon
dynamics~\cite{DW-1}.

By further carefully examining $\gamma(T)$ [see Fig. 3(d)], we show,
as expected that, a crossover from superdiffusive ($1<\gamma<2$) to
normal transport ($\gamma=2$) at about $T_c \simeq 0.1$ is likely to
take place, though longer space size simulations remain required to
confirm the transition. Surprisingly, these results also indicate a
continuous variation of $\gamma$ ($1.4< \gamma <2$), thus further
supporting that the universality classes of $\gamma$ may be changed
when the system's dispersion becomes so unusual.

Finally let us discuss why here we see a crossover between different
universality classes~\cite{Note-6}. As all the observed density
$\rho(x,t)$ show good coincidence with L\'{e}vy walks profiles, then
a natural picture in mind is that, phonons, as main heat carriers,
might perform various kinds of CTRW to support the different scaling
exponents, probably induced by many complicated environments. We
thus argue that such surroundings of phonons may mainly result from
the very peculiar phonon dispersions along with nonlinearity. Then
under appropriate anharmonicity and temperatures, many distinct
nonlinear excitations, such as solitary waves~\cite{Soliton},
discrete breathers (DBs)~\cite{DBs} and soft modes~\cite{Softs} will
be excited in various kinds of systems with quite different
dispersions. Such excitations will interact with phonons, eventually
leading to the different $\gamma$. With this picture we now turn to
analyzing the power spectrum of thermal fluctuations, which covers
both phonons and their environments information and may provide a
suggestive understanding of the observed crossover.
\begin{figure}
\centerline{\includegraphics[width=9.cm]{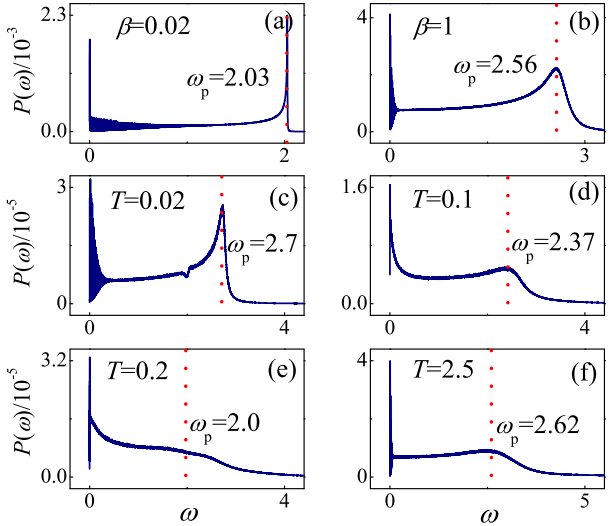}} \vskip-0.3cm
\caption{(Color online) $P(\omega)$ of the thermal fluctuations
($L=2000$): (a)-(b) FPU-$\beta$ chains; (c)-(f) DW chains. The
dotted lines indicate $\omega_{\rm{p}}$ in high frequency regime.}
\end{figure}

Figure 4 depicts the typical spectrum $P(\omega)$, defined by
$P(\omega)=\lim_{\tau \rightarrow \infty} \frac{1}{\tau}
\int_{0}^{\tau} v (t) \exp (- \rm{i} \mit{\omega} t) \rm{d} \mit{t}$
and calculated by a frequency $\omega$ analysis of the equilibrium
velocity $v(t)$ along the chains (for the calculation details, see
the appendix of the review~\cite{NianbeiLi_Review}). As expected,
$P(\omega)$ shows strong $\beta$ and $T$ dependence, similarly to
$\gamma$. We here address two points. First, in view of the whole
spectrum, as $\beta$ increases, FPU-$\beta$ system's $P(\omega)$
walks towards the direction of high frequencies, suggesting that
phonons tend to become ``harder" [see Fig. 4(a)-(b)]. Further
careful examination of the excitations indeed supports the emerging
of both solitary waves~\cite{Soliton} and DBs~\cite{DBs} (not
shown); rather, in the DW systems, phonons seem to become ``softer",
especially around $T \simeq 0.2$ [see Fig. 4(c)-(f)], indicating
that here the soft modes~\cite{Softs} appear. These tendencies can
be readily captured from the indicated peaks in the high frequencies
regime~\cite{Note-4}.
\begin{figure}
\centerline{\includegraphics[width=9.cm]{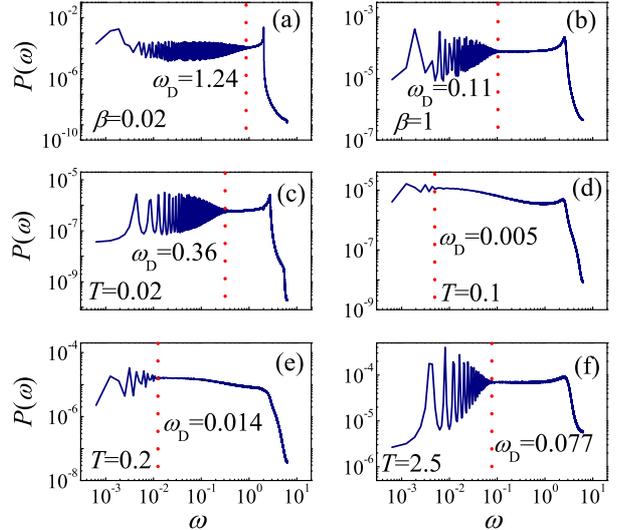}} \vskip-0.3cm
\caption{(Color online) Log-log plot of Fig. 4, where the dotted
lines indicate $\omega_{\rm{D}}$ below which phonons are damped
weakly. }
\end{figure}

We refer the second point to the lowest frequency components (see
Fig. 5). Phonons with the lowest frequency, usually called
long-wavelength (goldstone) modes, are generally believed to be very
weakly damped due to the conserved feature of
momentum~\cite{Newbook}. Because of their weak damping, the lowest
frequency modes can greatly affect heat transport. From Fig. 5 the
damping of phonons first originates from high frequencies and then
quickly walks towards the low ones, also showing strong $\beta$
($T$)-dependence. In particular, in the DW case, the strongest and
fastest damping seems to appear around $T_c \simeq 0.1$ [see Fig.
5(d)], which may explain why a transport close to normal could be
found.

For the FPU-$\beta$ chains, up to $\beta_c \simeq 0.01$-$0.02$, a
fast damping can be observed [see Fig. 5(a)-(b)]; however, we are
unable to identify $\beta_c$ just from the damping. Fortunately,
$\beta_c$ seems related to the strong stochasticity threshold
addressed in~\cite{New-2}. For the particular $\beta=1$ case,
Ref.~\cite{New-2} suggested an energy-density threshold about
$\langle E \rangle _c \simeq 0.1$. Then viewing that in our case
$\langle E \rangle \simeq 1.0$ (since $T=0.5$), a straightforward
analysis of $\langle E \rangle _c \sim \beta_c^{1/2}$ infers
$\beta_c \simeq 0.01$, which is consistent with the results of Fig.
2.

Given the above macroscopic and microscopic evidences, it would be
suggestive to propose such an understanding of the different
universality classes: \emph{different nonlinear spectrum dynamics
can result in quite distinct heat transport behaviors described by
phonons performing various kinds of CTRW, eventually leading to the
different scaling exponents}. We emphasize that here this picture
attributes the main mechanism to the phonons dispersions along with
nonlinearity, which seems not contradicted by the above direct
dynamics simulations.

In summary, by employing two representative oscillator systems
bearing (and not) usual phonon dispersions, we have demonstrated
with precise and convincing numerical evidences that, there is a
crossover between different universality classes of the space-time
scaling exponent $\gamma$ for heat transport in various kinds of 1D
systems with different phonon dispersions. With the difference of
phonons dispersions in mind and by varying the anharmonicity and
temperature, we have showed that $\gamma$ can be feasibly tuned from
$\gamma=1$ to $\gamma=2$ (for the two focused systems), indicating
that the corresponding transport can be controlled from ballistic,
superdiffusive ($1<\gamma<2$), to normal. We have also suggested an
understanding for the crossover by assuming phonons performing
various kinds of CTRW induced by the peculiar dispersions along with
nonlinearity, which seems to be supported by the analysis of both
systems phonon spectrum. These findings and the suggested picture
thus imply some potential applications. For example, in virtue of
the DW systems, one may be able to vary the phonons spectrum by
tuning system temperatures, and finally manipulate heat. Such an
idea would be realized by variation of the trapping frequencies in
the recent focused ion chains~\cite{Icon-1234}, where a structural
phase transition very similar to the DW systems has been
found~\cite{Icon-56}.

As we have understood this crossover, then apart from those
applications, further careful theoretical examinations, especially
the numerical ones, are expected to be stimulating. In fact, the
quite efficient direct dynamics simulation method~\cite{Method-2}
adopted here has special advantages for identifying the scaling of
transport, because it is applicable to any other physical quantities
diffusion in a large variety of complicated systems with any phonon
dispersion under various anharmonicity and temperatures; however,
many of them are not accessible by existing theories. Thus we expect
to extend this method to other kinds of systems and higher
dimensions (it is straightforward). For example, a quick application
to two-dimensional systems will be studied.
\begin{figure}
\centerline{\includegraphics[width=9.cm]{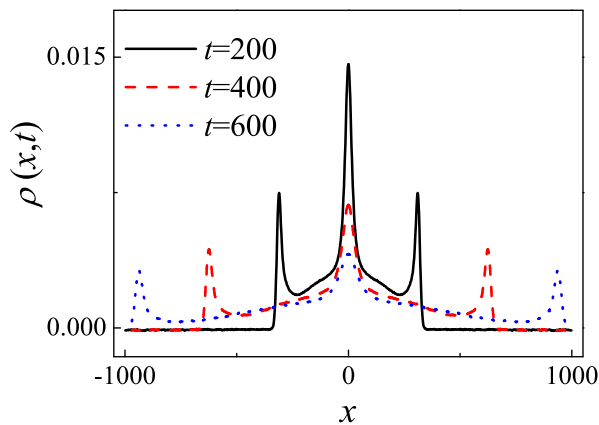}} \vskip-0.3cm
\caption{(Color online) $\rho(x,t)$ for the FPU-$\beta$ chain with
both NN and NNN couplings ($L_{\rm{effetive}}=2000$; $\beta =0.2$;
$T=0.5$; and the ratio $r=0.25$).}
\end{figure}

Another aspect of further works may be devoted to the underlying
picture. It has been well
proposed~\cite{Levy-1,Levy-2,Levy-3,Levy-4} that anomalous heat
transport in 1D systems can be understood by phonons performing
L\'{e}vy walks. Though all of the above simulations seem in good
agreement with the L\'{e}vy walks description, in our opinion, the
walks may not always be the L\'{e}vy type as the phonon dispersions
could be more complicated. In fact, our quite recent simulations of
a FPU-$\beta$ model chain with both nearest-neighbor (NN) and
next-nearest-neighbor (NNN) couplings having a phonon dispersion of
a high order form~\cite{MyWork} [the Hamiltonian is $H= \sum_{k}
^{L} p_{k} ^2 /2 +V (q_{k+1}-q_{k})+ r V (q_{k+2}-q_{k})$ with $r$
the ratio of the NNN to NN couplings and the potential $V (\xi) =
\xi^2/2 + \beta \xi^4 /4$; note that here the dispersion depends on
the ratio $r$], indeed support a more complicated density shape
which has not yet been covered by the existing progress of L\'{e}vy
walks theory~\cite{Levy-1} (see Fig. 6, the central ranges imply
some localizations, which may be due to the effects of the peculiar
dispersions along with nonlinearity enabling us to excite the
intraband DBs~\cite{MyWork}; note that here the L\'{e}vy walks
scaling might not be satisfied, so we are unable to show the exact
scaling exponent $\gamma$). That is why we attribute the mechanism
to generic CTRW but not it is limited to L\'{e}vy walks. We also
note that further efforts to understand the picture may have to
provide the random walks mechanism with a ``hydrodynamic"
foundation~\cite{Levy-1}. However, regardless of the picture, Fig. 6
again supports the fact that different universality classes will be
observed in different 1D nonlinear systems with quite different
phonon dispersions.

\acknowledgments The author is indebted to Profs. Hong Zhao, Jiao
Wang, and Yong Zhang for many valuable discussions, and appreciates
the kind encouragement from Prof. Sergey V. Dmitriev. This work was
supported by the NNSF of China (Grants No. 11575046 and No.
11205032)  and the NSF of Fujian province, China (Grant No.
2013J05008).


\begin{thebibliography}{0}

\bibitem{Anomalous-1}
  \Name{Alder B. J. \and Wainwright T. E.}
  \REVIEW{Phys. Rev. Lett.}{18}{1967}{988}.

\bibitem{Anomalous-2}
  \Name{Ernst M. H., Hauge E. H. \and van Leeuwen J. M. J.}
  \REVIEW{J. Stat. Phys.}{15}{1976}{7}.

\bibitem{Renormalize}
  \Name{Narayan O. \and Ramaswamy S.}
  \REVIEW{Phys. Rev. Lett.}{89}{2002}{200601}.

\bibitem{MCT-1}
  \Name{Delfini L., Lepri S., Livi R. \and Politi A.}
  \REVIEW{Phys. Rev. E}{73}{2006}{060201(R)}.

\bibitem{MCT-2}
  \Name{Delfini L., Lepri S., Livi R. \and Politi A.}
  \REVIEW{J. Stat. Mech.}{}{2007}{P02007}.

\bibitem{Lee-Dads-1}
  \Name{Lee-Dadswell G. R., Nickel B. G. \and Gray C. G.}
  \REVIEW{Phys. Rev. E}{72}{2005}{031202}.

\bibitem{Lee-Dads-2}
  \Name{Lee-Dadswell G. R., Nickel B. G. \and Gray C. G.}
  \REVIEW{J. Stat. Phys.}{132}{2008}{1}.

\bibitem{Nonuniversal_new}
  \Name{Lee-Dadswell G. R.}
  \REVIEW{Phys. Rev. E}{91}{2015}{032102}.

\bibitem{Hydrodynamics-1}
  \Name{vanBeijeren H.}
  \REVIEW{Phys. Rev. Lett.}{108}{2012}{180601}.

\bibitem{Hydrodynamics-2}
  \Name{Mendl C. B. \and Spohn H.}
  \REVIEW{Phys. Rev. Lett.}{111}{2013}{230601}.

\bibitem{Hydrodynamics-3}
  \Name{Spohn H.}
  \REVIEW{J. Stat. Phys.}{154}{2014}{1191}.

\bibitem{Levy-1}
  \Name{Zaburdaev V., Denisov S. \and Klafter J.}
  \REVIEW{Rev. Mod. Phys.}{87}{2015}{483}.

\bibitem{Levy-2}
\Name{Zaburdaev V., Denisov S. \and H\"{a}nggi P.}
  \REVIEW{Phys. Rev. Lett.}{106}{2011}{180601}.

\bibitem{Levy-3}
  \Name{Lepri S. \and Politi A.}
  \REVIEW{Phys. Rev. E}{83}{2011}{030107(R)}.

\bibitem{Levy-4}
  \Name{Dhar A., Satio K. \and Derrida B.}
  \REVIEW{Phys. Rev. E}{87}{2013}{010103(R)}.

\bibitem{Grassberger} \Name{Grassberger
P.,Nadler W., \and Yang L.}
\REVIEW{Phys. Rev.
Lett.}{89}{2002}{180601}.

\bibitem{Numerics-1}
  \Name{Lepri S., Livi R. \and Politi A.}
  \REVIEW{Phys. Rep.}{377}{2003}{1}.

\bibitem{Numerics-2}
  \Name{Dhar A.}
  \REVIEW{Adv. Phys.}{57}{2008}{457}.

\bibitem{Numerics-3}
  \Name{Mai T., Dhar A. \and Narayan O.}
  \REVIEW{Phys. Rev. Lett.}{98}{2007}{184301}.

\bibitem{Numerics-4}
  \Name{Denisov S., Klafter J. \and Urbakh M.}
  \REVIEW{Phys. Rev. Lett.}{91}{2003}{194301}.

\bibitem{Numerics-5}
  \Name{Cipriani P., Denisov S. \and Politi A.}
  \REVIEW{Phys. Rev. Lett.}{94}{2005}{244301}.


\bibitem{Numerics-6}
  \Name{Das S. G., Dhar A., Satio K., Mendl C. B. \and Spohn H.}
  \REVIEW{Phys. Rev. E}{90}{2014}{012124}.


\bibitem{Numerics-7}
  \Name{Mendl C. B. \and Spohn H.}
  \REVIEW{Phys. Rev. E}{90}{2014}{012147}.


\bibitem{Method-1}
  \Name{Zhao H.}
  \REVIEW{Phys. Rev. Lett.}{96}{2006}{140602}.


\bibitem{MyWork}
  \Name{Xiong D., Wang J. Zhang Y. \and Zhao H.}
  \REVIEW{Phys. Rev. E}{85}{2012}{020102(R)};
  \Name{Xiong D., Zhang Y. \and Zhao H.}
  \REVIEW{Phys. Rev. E}{88}{2013}{052128};
  \Name{Xiong D., Zhang Y. \and Zhao H.}
  \REVIEW{Phys. Rev. E}{90}{2014}{022117}.

\bibitem{Nonuniversal_new_1}
  \Name{Hurtado P. I. \and Garrido P. L.}
    arXiv:1506.03234v1.

\bibitem{Nonuniversal_new_2}
  \Name{Kosevich Y. A. \and Savin A. V.}
    arXiv:1509.03219v1.

\bibitem{Note} There may be a generic relationship between $\alpha$ and $\gamma$; such as $\alpha=2-\gamma$ suggested by the L\'{e}vy walks theory, obtained from simultaneously combining $\alpha=\mu-1$ and $\mu=3-\gamma$, where $\mu$ is the time scaling exponent of the mean squared displacement of energy fluctuations spreading $\sigma ^2 (t)$: $\sigma ^2 (t) \sim t^{\mu}$. However, it should be noted that $\alpha=2-\gamma$ has not been fully confirmed yet as the relation between $\alpha$ and $\mu$ is still controversial.

\bibitem{Rotor-1}
  \Name{Das S. G. \and Dhar A.}
    arXiv:1411.5247v2.

\bibitem{Rotor-2}
  \Name{Spohn H.}
    arXiv:1411.3907.

\bibitem{Workshop} The importance of the space-time scaling for thermal transport has been emphasized by S. Olla on a discussion session led by Lebowitz in a recent workshop (2012): \emph{Nonequilibrium statistical mechanics: mathematical understanding and numerical simulation}, see the final report of the workshop written by J. Lebowitz, S. Olla, and G.
Stoltz, see http://www.birs.ca/events/2012/5-day-workshops/12w5013.

\bibitem{KPZ}
  \Name{Kardar M., Parisi G. \and Zhang Y. C.}
  \REVIEW{Phys. Rev. Lett.}{56}{1986}{889}.


\bibitem{PRL2014}
  \Name{Popkov V., Schmidt J. \and Sch\"{u}tz G. M.}
  \REVIEW{Phys. Rev. Lett.}{112}{2014}{200602}.


\bibitem{2015New}
  \Name{Popkov V., Schadschneider A., Schmidt J. \and Sch\"{u}tz G.
  M.}
    \REVIEW{Proc. Natl. Acad. Sci.
    U.S.A.}{112}{2015}{41}.

\bibitem{Method-2}
  \Name{Cheng S., Zhang Y., Wang J. \and Zhao H.}
  \REVIEW{Phys. Rev. E}{87}{2013}{032153}.


\bibitem{Note-1} With $L$ unit mass particles. The averaged pressure is fixed to be zero, so
the number of particles is equal to the space size.

\bibitem{Liquid}
  \Name{Hansen J. P. \and McDonald I. R.}
  \Book{Theory of Simple Liquids}
  \Publ{Academic, London,}
  \Year{2006}.

\bibitem{New_shunda}
  \Name{Cheng S., Zhang Y., Wang J. \and Zhao H.}
  \REVIEW{Sci. China-Phys. Mech.
  Astron.}{56}{2013}{1466}.

\bibitem{New-Note-2} The scaling of the two side sound modes are also predicted by the related theories,
here we only limit our focus to the central heat mode scaling.

\bibitem{Note-3} This temperature has been verified to be quite accessible by
our simulations for FPU-$\beta$ systems.

\bibitem{Levy-Ballistic}
  \Name{Froemberg D., Schmiedeberg M., Barkai E. \and Zaburdaev V.}
  \REVIEW{Phys. Rev. E}{91}{2015}{022131}.

\bibitem{Barkai}
\Name{Margolin G. \and Barkai E.}
  \REVIEW{Phys. Rev. Lett.}{94}{2005}{080601};
    \Name{Stefani F. D., Hoogenboom J. P. \and Barkai E.}
  \REVIEW{Phys. Today}{62}{2009}{34}.

\bibitem{Note-5} Our preliminary results showed a variation of $\gamma$ with the increase of $\beta$, quite similar to the Fibonacci family of universality classes as suggested by~\cite{ Lee-Dads-1,Lee-Dads-2,Nonuniversal_new, 2015New}. In fact, the results of Fig. 2 in this work also have indicated some
hints.

\bibitem{New-Note} The lowest temperature considered here is $T=0.02$, up to which we have confirmed that the final results are not sensitive to the initial states assigned to just one of the potential wells; or between the two wells alternately, or randomly.

\bibitem{DW-234}
  \Name{Livi R., Politi A. \and Vassalli M.}
  \REVIEW{Phys. Rev. Lett.}{84}{2000}{2144};
  \Name{Gendelman O. V. \and Savin V.}
  \REVIEW{Phys. Rev. Lett.}{84}{2000}{2381};
  \Name{Roy D.}
  \REVIEW{Phys. Rev. E}{86}{2012}{041102}.
\Name{Li H.}
\REVIEW{Int. Mod. Phys. B}{25}{2011}{823}.

\bibitem{DW-1}
  \Name{Lee W., Kovacic G. \and Cai D.}
  \REVIEW{Proc. Natl. Acad. Sci. U.S.A.}{110}{2013}{3237}.

\bibitem{Note-6} The crossover means: there may be a particular universality classes existing under certain system parameter regimes, just as predicted by some related theorietical models; however, in view of the different phonon dispersions for different systems, the crossover between different universality calsses should appear.

\bibitem{Soliton}
  \Name{Kartashov Y. V., Malomed B. A. \and Torner L.}
  \REVIEW{Rev. Mod. Phys.}{83}{2011}{247}.

\bibitem{DBs}
  \Name{Flach S. \and Willis C. R.}
  \REVIEW{Phys. Rep.}{295}{1998}{181};
  \Name{Flach S. \and Gorbach A. V.}
  \REVIEW{Phys. Rep.}{467}{2008}{1};
  \Name{Dmitriev S. V., Chetverikov A. P. \and Velarde M. G.}
  \REVIEW{Phys. Status. Solidi B}{252}{2015}{1682}.

\bibitem{Softs}
  \Name{Burns G. \and Scott B. A.}
  \REVIEW{Phys. Rev. Lett.}{25}{1970}{167}.


\bibitem{NianbeiLi_Review}
  \Name{Li N., Ren J., Wang L., Zhang G., H\"{a}nggi P. \and Li B.}
  \REVIEW{Rev. Mod. Phys.}{84}{2012}{1045}.

\bibitem{Note-4} For $T=0.2$ [see Fig. 4(e)] the peak is displaced by
a platform with a very slight gradient, we just denote the center of
the platform.

\bibitem{Newbook}
  \Name{Pomeau Y. \and R\'{e}sibois R.}
  \REVIEW{Phys. Rep.}{19}{1975}{63}.


\bibitem{New-2}
  \Book{Anomalous transport: foundations and applications}
  \Editor{Klages R., Radons G. and Sokolov I. M.}
  \Publ{Wiley-VCH, Berlin}
  \Year{2006};
  \Name{Lepri S., Livi R. \and Politi A.}
  \Section{Anomalous heat conduction}.

\bibitem{Icon-1234}
  \Name{H\"{a}ffner H., Roos C. F. \and Blatt R.}
  \REVIEW{Phys. Rep.}{469}{2008}{155};
  \Name{Schneider C., Porras D. \and Schaetz T.}
  \REVIEW{Rep. Prog. Phys.}{75}{2012}{024401};
  \Name{Blatt R. \and Roos C. F.}
  \REVIEW{Nat. Phys.}{8}{2012}{277};
  \Name{Bermudez A., Bruderer M. \and Plenio M. B.}
  \REVIEW{Phys. Rev. Lett. }{111}{2013}{040601}.

\bibitem{Icon-56}
  \Name{Ruiz A., Alonso D., Plenio M. B. \and del Campo A.}
  \REVIEW{Phys. Rev. B}{89}{2014}{214305};
  \Name{Freitas N., Mart\'{\i}nez E. \and Pablo Paz J.}
    \REVIEW{Phys. Scr.}{91}{2015}{1};


\end{thebibliography}
\end{document}